\documentclass[12pt]{article}
\usepackage{graphicx}
\usepackage{amssymb}
\usepackage {amsmath}
\usepackage{latexsym}

\begin{document}

\noindent
\footnotesize{V. Krasnoholovets, A sub microscopic description of the diffraction phenomenon,   \\   
\textit{Nonlinear Optics and Quantum Optics} \textbf{41}, No. 4, 273-286 (2010)}

\begin{center}
\section*{A Sub Microscopic Description of   \\  the Diffraction Phenomenon}
\end{center}

\begin{center}
\bf {Volodymyr Krasnoholovets}
\end{center}

\vspace{2mm}

\begin{center}
Indra Scientific, Square du Solbosch 26, Brussels, B-1050, Belgium \\
E-mail: v{\_}kras@yahoo.com
\end{center}

\vspace{5mm}

\begin{abstract}
It is shown that a detailed sub microscopic consideration denies the
wave-particle duality for both material particles and field
particles, such as photons. In the case of particles, their $\psi
$-wave function is interpreted as the particle's field of inertia
and hence this field is characterised by its own field carriers,
inertons. Inertons and photons are considered as quasi-particles,
excitations of the real space constructed in the form of a
tessel-lattice. The diffraction of photons is explained as the
deflection of photons from their path owing to transverse flows of
inertons, which appear in the substance under consideration at the
decay of non-equilibrium phonons produced by transient photons.

\bigskip

\textbf{Key words:} wave-particle; photons; inertons; diffraction of
photons

\bigskip

\textbf{PACS:} 2.25.Fx Diffraction and scattering;
 42.50.Ct Quantum description of interaction of light and matter; related
experiments; 42.50.Xa Optical tests of quantum theory

\end{abstract}

\section*{1. Structure of the real space}

In conventional quantum mechanics an undetermined `wave-particle' is
further substituted by a package of superimposed monochromatic
abstract waves. It is this approximation that gives rise to the
inequality of the wave number $\Delta {\kern 1pt} k$ and the
position $\Delta {\kern 1pt} x$ of the package under consideration,
which then results in Heisenberg's uncertainties [$\Delta {\kern
1pt} x$, $\Delta {\kern 1pt} p$]$ > h$, and related to de Broglie $k
= p/\lambda _{\rm {\kern 1pt} de {\kern 1pt} Br.} $.

In a simple way
Boyd [1] showed that photons are not subjects of Heisenberg
uncertainty; Boyd also referred to Hans G. Dehmelt who won the Nobel
Prize 1989 for the development of the ion trap technique
experiments. Dehmelt [2-5] proved that both the position and
momentum of an electron could be measured simultaneously; he kept a
practically motionless electron in an electromagnetic confinement
system for months, which allowed his team to measure simultaneously
- with accuracy $10^{ - 11}$ to $10^{ - 16}$ - the position,
momentum and other parameters.

Nevertheless, a wave-particle duality and the uncertainty principle
still remain significant in the quantum mechanical formalism. The
formalism was developed in an abstract phase space and the high end
of its applications is the size of the atom $\sim 10^{ - 10}$ m.
Quantum mechanics operates with canonical particles but does not
determine their origin nor an actual size. In quantum physics the
physical space is treated as an ``arena of action''. In such a
determination there exists: 1) subjectivity and 2) objects
themselves, which play in processes and can not be examined at all
(for instance, size, shape and the inner dynamics of the electron;
what is a photon?; what are the particle's de Broglie wavelength
$\lambda _{{\kern 1pt} \rm de {\kern 1pt} {\kern 1pt} Br.} $ and
Compton wavelength $\lambda_{{\kern 1pt}\rm Com.}$?; how to
understand the notion/phenomenon ``wave-particle''?; what is spin?;
what is the mechanism that forms Newton's gravitational potential
$G{\kern 1pt} m/r$ around an object with mass $m$ ?; what does the
notion `mass' mean exactly?, etc.). 

A few years ago a detailed theory of the real physical space was 
created by Michel Bounias and the author [6-9]. 
Initially the generalisation of the concept of
mathematical space was proposed, which was done through set theory,
topology and fractal geometry. This in turn allowed us to look at
the problem of the constitution of physical space from the most
fundamental standpoint. A physical space is derived from the
mathematical space that in turn is constructed as a mathematical
lattice of topological balls. This lattice of balls has been
referred to as a \textbf{\textit{tessel-lattice}}, in which balls
are found in a degenerate state and their characteristics are such
mathematical parameters as length, surface, volume and fractality.
The size of a ball in the tessel-lattice was associated with the
Planck's size $l_{{\kern 1pt}\rm  P} = \sqrt {\hbar G/c^{3}} \sim
10^{ - 35}$ m. Evidently, the removal of degeneracy must result in
local phase transitions in the tessel-lattice, which creates
``solid'' physical matter. So matter (mass, charge and canonical
particle) is immediately generated by space and has to be described
by the same characteristics as the balls from which matter is
formed. The behaviour of a canonical particle obeys submicroscopic
mechanics (see, e.g. review article [10]) that is determined on the
Planck's scale in the real space and is wholly deterministic by its
nature. At the same time, it has been shown that deterministic
submicroscopic mechanics is in complete agreement with the results
predicted by conventional probabilistic quantum mechanics, which is
developed on the atomic scale in an abstract phase space. Moreover,
submicroscopic mechanics allows the derivation of ..Newton's law of
universal gravitation and the ..nuclear forces starting from first
sub microscopic principles of the tessellation structure of physical
space\textbf{.} A particle appears as a local fractal volumetric
deformation in the tessel-lattice, i.e. a fractal volumetric
deformation of a cell of the tessel-lattice. The main peculiarity of
the theory is the availability of excitations of the tessel-lattice
around a moving particle. These excitations transfer fragments of
the particle's mass and are responsible for inertial properties of
the particle. Because of that they were called
\textbf{\textit{inertons.}} The following relationship was derived

\begin{equation} \Lambda = \lambda _{\kern 1pt \rm de{\kern 1pt} {\kern 1pt} Br.}
{\kern 1pt} {\kern 1pt} c/\upsilon \label{eq1}
\end{equation}

\noindent where $\lambda _{\rm {\kern 1pt} de{\kern 1pt} {\kern 1pt}
Br.} $ is the spatial amplitude/period of the particle associated
with the particle's de Broglie wavelength; $c$ and $\upsilon $ are
the velocity of light and the particle, respectively. The value of
$\Lambda $ in expression (1) determines the amplitude of the
particle's inerton cloud, which spreads in transversal directions
around the particle; along the particle's path it spreads up to the
distance $\lambda _{\rm {\kern 1pt} de{\kern 1pt} {\kern 1pt} Br.}
/2$. The volume around the particle occupied by its inertons has to
be treated as the field of inertia of the particle. Then the quantum
mechanical wave function $\psi $ becomes determined just in this
range and, therefore, the $\psi $-wave function represents an image
of the original field of inertia (i.e. particle's inerton cloud)
defined in the real space. The introduction of inertons makes the
principle of uncertainty superfluous, because in the real space
instead of an undetermined wave-particle we have two subsystems: the
particulate cell (the particle kernel) and the inerton cloud that
accompanies it. So far physicists have examined the behaviour of
only a bare, or unclosed particle, but the other subsystem, the
particle's inerton cloud, went unnoticed and has not been
considered. In submicroscopic mechanics, the uncertainty principle
has no relevance. Nevertheless, at measurements, the particle's
inerton cloud is strongly scattered, i.e. the particle looses its
inerton cloud, which immediately prescribes a probability to its
behaviour. The present study shows that including the particle's
inerton cloud is important for examination of subtle kinetics of
processes pertaining to the interaction of photons with
non-polarisable matter and the diffraction of photons. Besides,
inertons manifest themselves at photon-photon crossing.

\begin{figure}
\includegraphics[width=5.5in]{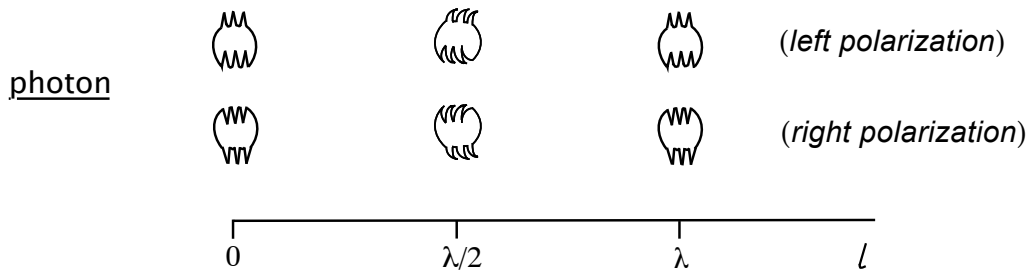}
\caption{Structure of the photon. The electrical polarisation, when
needles are normal to the spherical surface, appears with the
interval of $\lambda $. Needles are periodically combed, which
physically means the appearance of the magnetic field in the present
point. If needles are combed towards the direction of motion of the
photon, the photon can be called right-polarised. If needles are
combed in the reverse direction of the motion of the photon, the
photon can be called left-polarised.}
\end{figure}

\section*{2. Structure of the photon}

The inerton is a basic excitation of the real space, which transfers
fragments of mass (i.e. local deformation of a cell) and fractality.
The photon is the second basic excitation of the space.

The photon appears [11-13] as a polarisation state of the surface of
the inerton. These two fundamental quasi-particles of space can
exist only in the state of motion. We can draw the appropriate
picture of the photon as follows: the mass (local deformation) of
the migrating photon oscillates, periodically transforming to the
state that can be described as the tension of the cell. The geometry
of the surface of the photon oscillates between the state of normal
needles (electric polarisation) and the state of combed needles
(magnetic polarisation).

Since we compare the size of an elementary cell of the
tessel-lattice with the Planck's fundamental length $l_{\rm {\kern
1pt} P}$, we shall attribute this scale as the actual size of the
photon. However, high-energy physics extrapolates the unification of
three types of interactions (electromagnetic, weak and strong) on
the scale $\sim 10^{-30}$ m. This would mean that although the core
of the photon occupies only one cell, a certain fluctuation in the
tessel-lattice may reach up to the scale $10^{-30 }$ m.

Figure 1 represents an instantaneous photo of the photon\textbf{:}
it is a cell of the tessel-lattice whose upper part of the surface
is covered by needles that stick out of the cell and the lower part
of the surface is covered by needles that stick inside of the cell.
Owing to certain non-adiabatic processes, for example, a collision
of the charged particle's photon cloud with an obstacle, free
photons are released from the photon cloud that surrounds the
charged particle.

A free photon migrates in the tessel-lattice by hopping from cell to
cell. During such a motion the state of its surface periodically
changes between the state of normal needles (electric polarisation)
and the state of combed needles (magnetic polarisation). The photon
in each odd section $\lambda /2$ of its path looses the electric
polarisation, which is going to zero, and acquires the magnetic
polarisation; in even sections $\lambda /2$ of the photon's path it
looses the magnetic polarisation but restores its electric
polarisation. Thus the wavelength $\lambda $ of the photon
represents a spatial period in which the polarisation of the photon
is transformed from pure electric to pure magnetic. Having $\lambda
$ and knowing the velocity $c$ of a free photon we can calculate the
photon frequency, which features the frequency of transformation of
magnetic and electric polarisations: $\nu = c/\lambda $.

\section*{3. Quantum theory of diffraction}

Epstein and Ehrenfest [14,15] following Compton considered a three
dimensional infinite triclinic lattice with the spacings $a_{x} $,
$a_{y} $ and $a_{z} $ in the respective directions of its chief
axes. They believed that in a collision with a light quantum such a
lattice could only pick up a linear momentum the orthogonal
projections of which $p_{x} $, $p_{y} $ and $p_{z} $ on the
directions $x$, $y$ and $z$ of the chief axes satisfy the
fundamental conditions of the quantum theory

\begin{equation}
\int {p_{x}}  dx = n_{x} h,\;\quad \int {p_{y}}  dx = n_{y} h,\quad
\;\int {p_{z}}  dx = n_{z} h  \label{eq2}
\end{equation}

\noindent here $n_{i} $ are three integral numbers and $h$ denotes
Planck's constant of action. The periodicity of the lattice is given
by its spacings $a_{i} $ so that the first integral is to be
extended from $x$ to $x + a_{x} $ and the others correspondingly.
This allowed them to obtain

\begin{equation}
p_{x} = h{\kern 1pt} {\kern 1pt} n_{x} /a_{x}, \quad \; p_{x} =
h{\kern 1pt} {\kern 1pt} n_{x} /a_{x} ,\quad \;p_{x} = h{\kern 1pt}
{\kern 1pt} n_{x} /a_{x}. \label{eq3}
\end{equation}

Then they compared relationships (3) to relations for light, because
the momentum of a light quantum (i.e. photon) of the frequency $\nu
$ is given by $h\nu /c = h/\lambda $, where $\lambda $ is the
wavelength in vacuum corresponding to the frequency $\nu $. The
principle of conservation of momentum requires the relations

\begin{equation}
\alpha - \alpha _{0} = \lambda {\kern 1pt} n_{y} /a_{y} ,\; \;\beta
- \beta _{0} = \lambda  {\kern 1pt} n_{z} /a_{z} ,\; \;\gamma -
\gamma _{0} = \lambda {\kern 1pt} {\kern 1pt} n_{z} /a_{z},
\label{eq4}
\end{equation}

\noindent where $\alpha _{0} ,\;{\kern 1pt} \beta _{0} ,\;{\kern
1pt} \gamma _{0} $ and $\alpha ,\;{\kern 1pt}  \beta ,{\kern 1pt}
\,\gamma $ are cosines between main axes respectively before and
after collisions with sites of the lattice. These relationships are
identical with those derived by von Laue from the theory of
interference.

Epstein and Ehrenfest mention that the distribution of electronic
density is sinusoidal in the lattice and hence can be presented by
the formula

\begin{equation}
\rho = A{\kern 1pt} {\kern 1pt} \sin{\kern 1pt} \left( {2{\kern 1pt}
\pi {\kern 1pt} x/a_{x} + \delta}  \right) \label{eq5}
\end{equation}

\noindent $\rho $ in an infinite grating is

\begin{equation}
\rho = \sum\limits_{n = 0}^{\infty}  {A_{n}}  \sin\left( {2{\kern
1pt} \pi n{\kern 1pt} x + \delta}  \right) \label{eq6}
\end{equation}

They further said that following the Fourier theorem any
distribution of electronic density could be built up of sinusoidal
terms, i.e. could be presented as a superposition of infinite
sinusoidal gratings of the type (6).

Ehrenfest and Epstein [14,15] note that some kinds of diffraction,
e.g. the Fresnel ones, could not be explained by purely corpuscular
considerations and essential features of the wave theory in a form
suitable for the quantum theory would be needed. They believed that
quanta of light should attribute phase and coherence similar to the
waves of the classical theory. And they assumed the first papers by
de Broglie and Schr\"odinger on modern quantum mechanics would bring
researchers much nearer to the solution of the problem…

The problem was resolved by introducing an undetermined notion of
``wave-particle'', though Louis de Broglie, the ``father'' of
quantum relationships $E = h\nu $ and $\lambda _{\rm {\kern 1pt}
de{\kern 1.5pt} Br.} = h/( m \upsilon )$ for a particle was against
such unification. Nevertheless, by using this strange ``monster''
called the wave-particle duality, physicists were able to explain
some previously unknown phenomena.

Panarella [16] wrote a remarkable review paper dedicated to the
experimental testing of the wave-particle duality notion for
photons. He reviewed the results of many researchers and also
presented his own data and the analysis. In particular, he
emphasized that his experimental results brought new evidence that a
diffraction pattern on a photographic plate is not presented when
the intensity of light was extremely low, even when the total number
of photons reaching the film is larger than that which was needed to
form a clear diffraction pattern. Some of his experiments lasted for
weeks! Thus it was established that a diffraction pattern did not
follow the linear principle with decreasing light intensity, as the
wave-particle duality required. He obtained the same results by
using photoelectric detection and oscilloscope recording of the
diffraction pattern.

In particular, Panarella [16] notes that with a flux (generated by
an optical laser) of around $10^{10}$ statistically independent
photons/sec in the interferometer, a clear diffraction pattern is
recorded on the oscilloscope. At a photon flux of around $10^{8}$
photons/sec, no clear diffraction pattern appears. The further
decrease of the intensity shows an increase of nonlinearity in the
behaviour of photons. Moreover, a flux in the interferometer of
$10^{4}$ photons/sec shows that we deal with a single particle
phenomenon - no diffraction at all. Analysing the experiments of
previous researchers who dealt with fluxes of only tens of photons
per second, Panarella rightly intimated that they were unable
unambiguously to determine whether their sources of light produced
individual/single photons or the sources produced packets of
photons.

Panarella concludes: ``The series of experiments reported here on
the detection of diffraction patterns from a laser source at
different low light intensities confirms the wave nature of
collections of photons but tends to dispute it, or not provide a
clear proof of it, for single photons''.

Further on, Panarella [16] tries to develop a ``photon clump'' model
in which he hypothesises a possible interaction between single
photons in a low intensity photon flux, which gathers photons in
clumps, such that they do not show wave properties at the
diffraction. However, his hypothesis raises the serious problem of
the inner nature of such interaction (sub-electromagnetic
interaction between photons?).

\section*{4. Inertons as the reason for the diffraction phenomenon}

Since before reaching the target photons pass through the
interferometer, which includes a series of details (lenses, mirrors,
etc. and a foil(s) with a pinhole), we have to concentrate on some
of its peculiarities, because they cause the photons to interfere.
In a transparent substance photons scatter by the structural
non-homogeneities producing non-equilibrium acoustic excitations
with wave numbers $k$ close to those of photons. If $\omega $ is the
cyclic frequency of an incident photon then the cyclic frequency of
the acoustic excitation (phonon) is [17]

\begin{equation}
\Omega \cong \frac{{2{\kern 1pt} \upsilon _{\rm sound} {\kern 1pt}
\omega {\kern 1pt} {\kern 1pt} n}}{c} \sin\frac{\varphi }{2} = 4\pi
{\kern 1pt} \frac{\kern 1pt \upsilon _{\rm sound} {\kern 1pt} {\kern
1pt} n}{\lambda } \sin\frac{\varphi}{2} \label{eq7}
\end{equation}

\noindent where $\lambda $ is the wavelength of the photon,
$\upsilon _{\rm sound} $ is the sound velocity of the substance and
$n$ its refraction index. $\varphi $ is the angle between the
initial and scattered photons, which can be treated as very small
for glass, $\varphi < < 1$, and hence the direction of motion of a
produced acoustic phonon is practically parallel to that of the
photon. The lifetime of generated acoustic excitations $\tau $ is
about $10^{ - 11}$ s in a metal [18] and $10^{ - 10}$ to $10^{ - 8}$
s in semiconductors and dielectrics [19-22]. This means that in a
short time $\tau $, non-equilibrium phonons decay. These
non-equilibrium phonons are the major subject of our study. In line
with our recent research [23], entities in condensed media behave
similar to single particles, namely, vibrating near equilibrium
positions they create clouds of inertons that accompany the
entities. That is, the amplitude of a vibrating atom in a solid is
considered as the atom's de Broglie wavelength. Therefore, we can
apply submicroscopic mechanics developed for free particles to
vibrating atoms as well. This means that in a solid we may use
expression (1) not only for atoms but also for phonons. Hence in the
background of the inerton field of equilibrium phonons, which can be
considered as noise, non-equilibrium phonons produced by incident
photons have to generate inertons in addition to the noise. During a
short time, non-equilibrium phonons gradually release generated
inertons in transverse directions to the phonon's wave vector $k$.
This means that these inertons move almost perpendicular to the beam
of photons and hence can tangibly affect the photon trajectories.
Pictures below demonstrate how forward photons generate - through
non-equilibrium phonons - flows of inertons in a transparent
substance, which then affect the subsequent photons of the same beam
of incident photons. We may assume that photons in a beam form a
three dimensional grid. Let the cross-section area of the laser beam
be $\pi {\kern 1pt} r^{2}$ where $r$ is the radius of beam. Then the
volume of photons per second in the beam, is $c{\kern 1pt} \pi
{\kern 1pt} r^{2}$. Therefore, the concentration of photons per
second is $N/( {c\pi {\kern 1pt} r^{2}})$ where $N$ is the number of
photons in a photon flux that passes the interferometer per second.
Having the concentration, we can derive the mean distance between
photons in the beam, $l = \left( {c\pi {\kern 1pt} r^{2}/N}
\right)^{ 1/3}$. A photon can travel this distance in a time $t =
l/c$. We may estimate this time $t$ for Panarella's experiments [16]
and compare it with the mentioned values of the relaxation time
$\tau $ of phonons in different media.

Why is it interesting to compare $t$ and $\tau $? Because in a
photon flux forward photons, which generate the emission of inertons
in the interferometer, are able to affect following photons by means
of the emitted inertons. The pictures below clearly demonstrate this
mechanism.

\begin{figure}
\begin{center}
\includegraphics[width=1.8in]{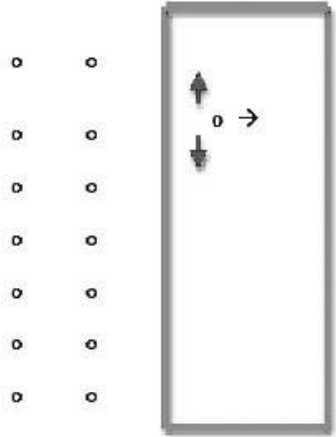}
\caption{ The first photon enters the interferometer. The photon
creates the acoustical excitation that in turn generates its cloud
of inertons in transverse directions (non-relevant photons are shown
before the interferometer).} \label{Figure 2}
\end{center}
\end{figure}

\begin{figure}
\begin{center}
\includegraphics[width=1.8in]{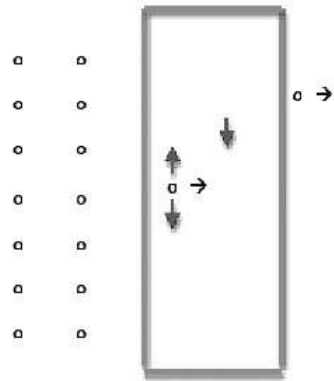}
\caption{ The first photon leaves the interferometer. The following
photon just entered the interferometer; the photon creates the
appropriate acoustic excitation, which generates a cloud of
inertons, and inertons generated by the previous photon are
approaching the path of the second photon.} \label{Figure 3}
\end{center}
\end{figure}

\begin{figure}
\begin{center}
\includegraphics[width=2.5in]{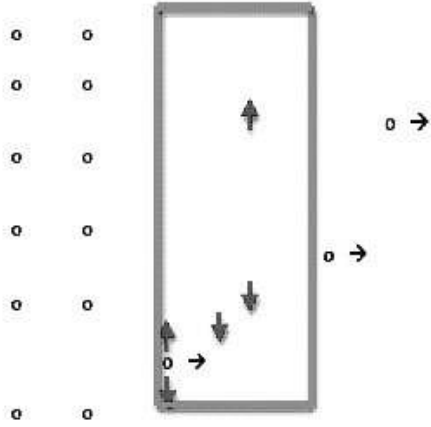}
\caption{The two photons have already left the interferometer: the
second one has experienced a sideways action through the inertons of
the first photon. The third photon just enters the interferometer;
it experiences sideways action through inertons generated by the two
previous photons (through the respective decayed phonons).
 } \label{Figure 4}
\end{center}
\end{figure}

\begin{figure}
\begin{center}
\includegraphics[width=3.3in]{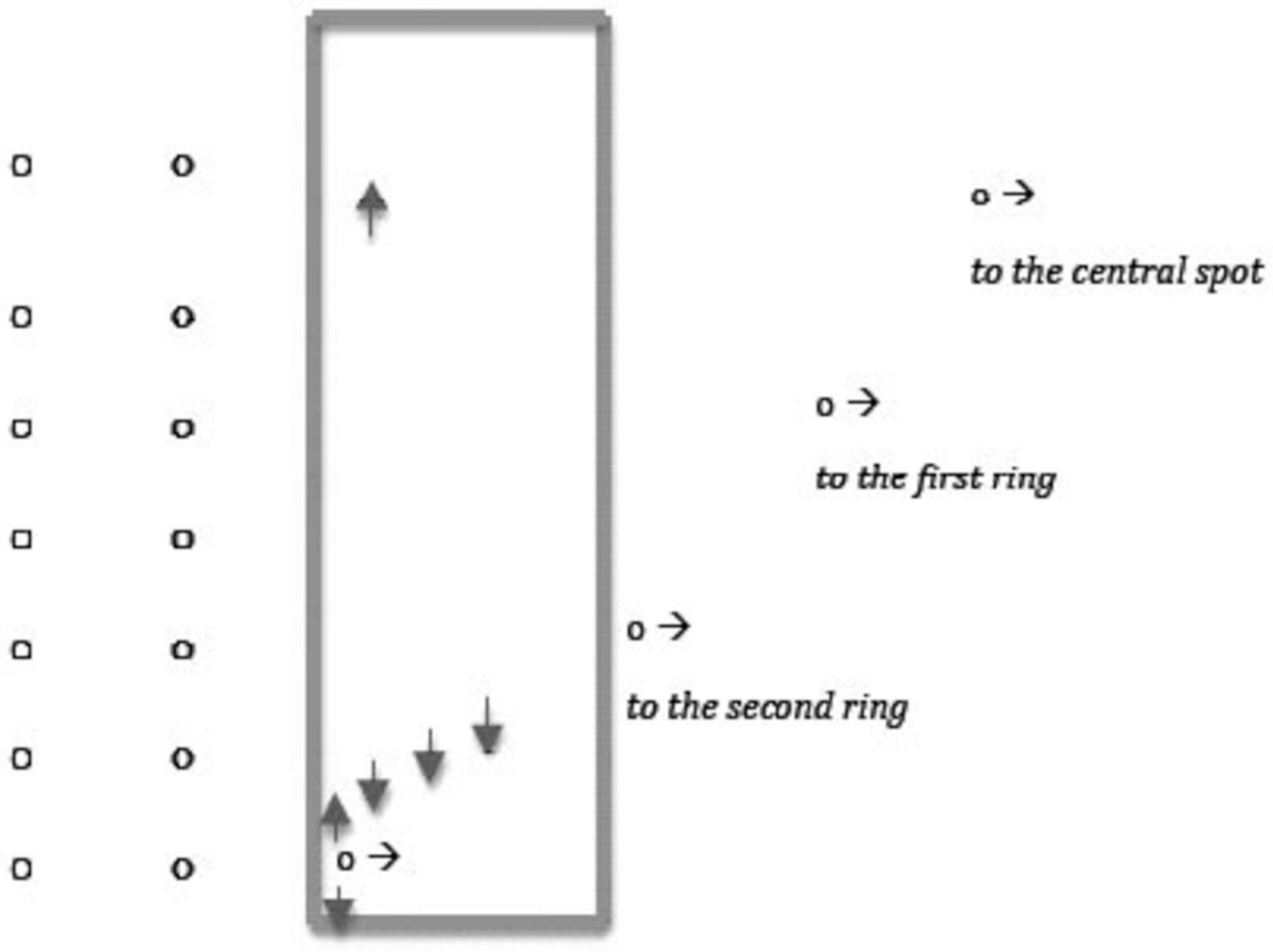}
\caption{The three photons have already left the interferometer and
the fourth photon that has entered the interferometer undergoes
sideways action from three flows of inertons generated by the
previous photons. The three first photons follow their own
trajectories: 1) the first one, which has not been affected by
inertons, follows to the centre of the target; 2) the second photon,
which was influenced by the first photon (through inertons of the
appropriate phonon), is going to form the first ring of the Airy
diffraction pattern; 3) the third photon, which underwent the
influence of the double flow of inertons (from the two first
photons), is deflected to forming the third ring of the Airy
pattern, and so on…} \label{Figure 5}
\end{center}
\end{figure}

A similar situation takes place in a foil at the edge of a pinhole.
Photons bombard the foil and generate non-equilibrium phonons. The
wave vector of phonons ${k}'$ practically coincides with the wave
vector of incident photons $k$. That is why the phonons decaying in
time $\tau $ generate inertons in transverse directions. These
inertons intersect the photon flux in the pinhole and are able to
affect photons there.

\begin{figure}
\begin{center}
\includegraphics[width=3.7in]{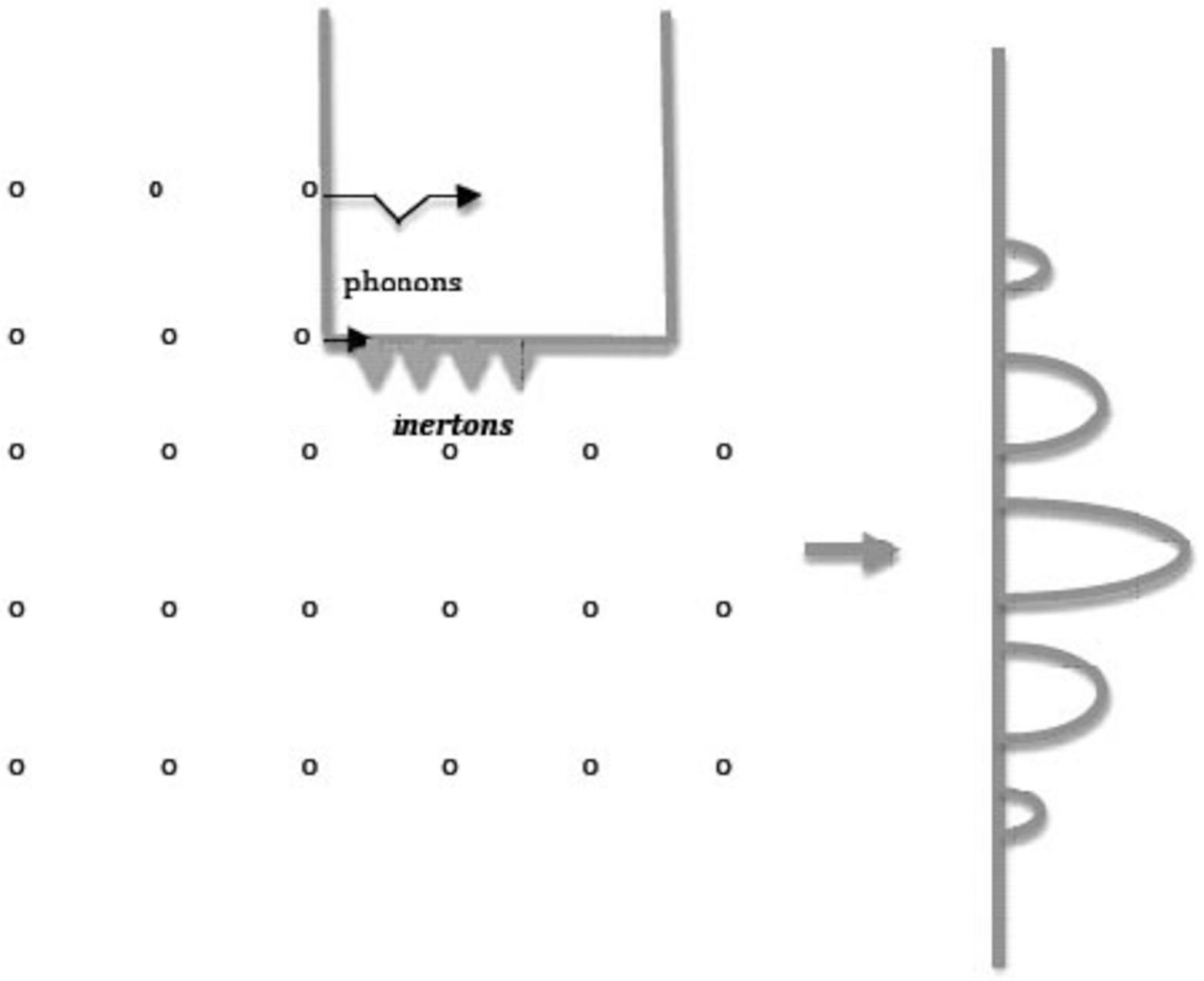}
\caption{Inertons intersect the pinhole affecting the flow of
photons, which results in the formation of the diffraction pattern -
the central peak with subsidiary maxima - on the target.}
\label{Figure 6}
\end{center}
\end{figure}

Let us estimate the value of $t$, i.e.
\begin{equation}
t = \frac{{\left( {c\pi {\kern 1pt} r^{2}/N} \right)^{ 1/3}}}{{c}},
\label{eq8}
\end{equation}

\noindent the time interval when a photon, which follows the
previous one, will arrive at the zone of action of inertons
generated by the forward photon through the production and decay of
a non-equilibrium phonon. Let the radius of the laser beam be $r
\approx $ 0.35 cm, then for the three sequential values of photon
intensities, used by Panarella [16], $N_{1} \approx 10^{10}$, $N_{2}
\approx 10^{8}$ and $N_{3} \approx 10^{4}$ we obtain from expression
(8): $t_{1} \approx 1.2 \times 10^{ - 10}$ s, $t_{2} \approx 6
\times 10^{ - 10}$ s and $t_{3} \approx 1.2 \times 10^{ - 8}$ s. The
lifetime of non-equilibrium phonons for dielectrics, as mentioned
above, varies from 10$^{-10}$ s to 10$^{-8}$ s [19-22]. Thus if the
inequality

\begin{equation}
t \ge \tau \label{eq9}
\end{equation}

\noindent holds, the second photon will arrive to the interferometer
at the moment when inertons generated by the first photon will
already be absent there. Therefore, the second photon does not
experience a transverse action and will continue to follow its path
to the central peak on the target. The inequality (9) holds for the
case of the lowest intensity of photons, $N_{3} \approx 10^{4}$
photons/sec, namely, $t_{3} > \tau $. Hence the mechanism described
is capable to account for Panarella's experiments in which the
diffraction fringe was absent.

The distribution of photons by rings of the diffraction pattern is
described in classical optics [24]: the first subsidiary maximum
should have an amplitude 0.0175 times the amplitude of the central
peak; the second subsidiary maximum has an amplitude 0.0042 times
the central amplitude. These results point out that the intensity of
transverse inerton flows in the interferometer, which deflects
photons from their direct way to the central peak, is not negligible
in the case of a comparative high intensity $N$ of the photon flux.
What is the reason for such perceptible intensity of inertons?

If the energy of an incident photon is $h\nu $, then the energy of
the acoustic excitation produced by the photon is $\hbar \Omega
\,\,\, \approx \,\,h\nu {\kern 1pt} {\kern 1pt} {\kern 1pt} \cdot
\upsilon _{\rm sound} /c\, \approx \,\,\,10^{ - 5}{\kern 1pt} h\nu
$. The energy $\hbar {\kern 1pt} \Omega $ is quenched during the
time $\tau $ and inertons emitted at the phonon decay carry away an
energy no more than $10^{ - 5}{\kern 1pt} h\nu $. This value of
energy is not enough to deflect a subsequent photon from the direct
line; this would simply fuzzify the width of the central spot from
the diameter $d_{{\kern 1pt} 0} $ to $\left( {1 + 10^{ - 5}}
\right){\kern 1pt} \cdot d_{{\kern 1pt} 0} $.

However, in the interferometer the initial photon produces hundreds
or even thousands of acoustic excitations $n_{\rm phon.} $ and hence
the intensity of the emitted inerton field will also be a thousand
times $10^{ - 5}{\kern 1pt} h\nu $. Then the position of the first
ring on the target will be determined by the expression

\begin{equation}
d_{{\kern 1pt} 1} = l \cdot \tan\frac{{\hbar {\kern 1pt} {\kern 1pt}
\Omega {\kern 1pt} {\kern 1pt} n_{\rm phon.}} }{{h\nu} } \approx
10^{ - 5}n_{\rm phon.} \cdot l \label{eq10}
\end{equation}

\noindent where $l$ is the efficient length of the interferometer;
the angle of deflection $\phi $ of photons (with the energy $h\nu $)
caused by an inerton flow generated by $n_{phon.} $ phonons (with
the energy $\hbar \Omega $) is given by the function $\tan\left[
{\hbar {\kern 1pt} {\kern 1pt} \Omega {\kern 1pt} {\kern 1pt} n_{\rm
phon.} /\left( {h\nu}  \right)} \right]$. In expression (10) we put
$\phi < < 1$, however, at the same time the flow of inertons is
still treated rather intensively, such that $d_{{\kern 1pt} 1} -
d_{{\kern 1pt} 0} > d_{{\kern 1pt} 0} $, i.e. the position of the
first ring does not overlap with the central spot. Then the second
ring is formed by a flow of inertons generated by $2{\kern 1pt}
n_{\rm phon.} $ phonons (after the first and the second photons),
the third ring is formed by inertons generated by $3{\kern 1pt}
n_{\rm phon.} $ (after the first, second and third photons), etc.

\section*{5. Concluding remarks}

We have analysed the kinetics of a photon flux in an interferometer.
The kinetics show that incident photons producing acoustic
excitations (phonons) are responsible also for the emission of
inertons. These inertons emerge at the decay of non-equilibrium
phonons in a short lifetime $\tau $ and spread in transverse
directions to the photon flux. The flow of inertons influences
subsequent photons of the photon flux, which deflects some photons
from the initial strait line. Following new trajectories, the
photons form subsidiary maxima around the central maximum on the
target.

It seems the parameter $n_{\rm phon.} $ (the number of phonons
needed to generate the transverse flows of inertons for the
deflection of photons) allows an experimental verification. Namely,
the classical diffraction pattern may appear only when in the
interferometer (for instance, a lens) the intensity of photons $N >
10^{4}$ photons/sec and the thickness exceeds some critical value.
Only starting from a concrete thickness of the lens the number of
acoustical excitations will be above the critical value, $n_{\rm
phon.}
> n_{\rm phon.}^{\left( {c} \right)} $, and only at this moment the
classical diffraction pattern will be able to emerge following the
mechanism described above.

Recently Cardone, Mignany and colleagues [25-27] have revealed
anomalous behaviour of photons at crossing photon beam experiments
in both the optical and the microwave range. They concluded that the
probability wave should be replaced by admitting an interpretation
in terms of the Einstein-de Broglie-Bohm ``hollow'' wave for
photons. Those experiments sustain the interpretation of the hollow
wave as a deformation of the space-time geometry. These experiments
further support the sub-microscopic concept, which has been applied
in this study for the explanation of diffraction and non-diffraction
of photons. Indeed, the crossing of photon beams has to result in a
partial annihilation of colliding photons, such that the surface
polarisation is eliminated from these field quasi-particles and only
a local volumetric fractal deformation remains. In other words, in
the photon-photon collisions the electromagnetic polarisation is
compensated and naked inertons appear instead of photons (recall,
the photon state is a state of the structured surface of an inerton;
the photon state appears on an inerton at the induction of the
surface fractality, as shown in Figure 1).

The author thanks greatly J. Perina (Jr.), O. Haderka, M. Dusek and J. Fiurasek, the organisers of the 11th International Conference on Squeezed States and  Uncertainty Relations (Olomouc, Czech Republic, 22-26 June 2009) at which this work was delivered.

\hspace{0.5cm}

\section*{AFTERWORD}

\hspace*{\parindent} 
After the publication of this paper I learned about the following papers that dealt 
with testing of the diffraction phenomenon:

\hspace{0.3cm}

\noindent
[A1] S. Jeffers, R. Wadlinger and G. Hunter, Low-light-level diffraction experiments: 
No evidence for anomalous effects,
\textit{Canad. J. Phys.}  \textbf{6}, 91471-1475 (1991).

\hspace{0.3cm}

\noindent
[A2] S. Jeffers and J. Sloan, A low light level diffraction experiment for anomalies research, 
\textit{J. Scientific Exploration} \textbf{6}, No. 4, 333-352 (1992).

\hspace{0.3cm}

\noindent [A3] Yu. P. Dontsov and A. I. Baz, Interference experiments with statistically independent photons, 
\textit{JETF} (\textit{\footnotesize Journal of Experimental and Theoretical Physics}) \textbf{52}, No. 1, 3-11 (1967); in Russian.  

\hspace{0.5cm}

Jeffers et al. [A1, A2] tried to repeat the Panarella's experiments [16]. Jeffers et al. reported a similar series of Panarella's low-intensity diffraction experiments (using two different optoelectronic detectors); the lowest intensity of a photon flux reached by Jeffers et al. was the same as was the case of Panarella [16], i.e. 10$^4$ photons/sec. However, their results did not substantiate the anomalous effects revealed by Panarella [16] and also Dontsov and Baz [A3].

It should be noted that Jeffers et al. [A1, A2] used the other kind of a hole than the hole used by Panarella; namely, they used a slit, which was bigger in size that a pinhole of Panarella [16]. This means that an area of the screen attacked by photons was also larger in the Jeffers' experiments. In other words, at Jeffers' conditions photons, which impacted the screen in the vicinity of the slit, launched non-stationary phonons even at the flux of 10$^4$ photons/sec; therefore the decay of excited non-stationary phonons produced inertons in transversal directions. Thus, flows of inertons after the decay of non-stationary phonons were constantly present inside the Jeffers' slit. Perhaps for the geometry used by Jeffers et al. the intensity of photon flux should be lower than 10$^4$ to obtain the result similar to Panarella [16] and Dontsov and Baz [A3].

Dontsov and Baz [A3] were able to achieve an extremely low intensity of statistically independent photons, 200 photons/sec. Their hole had the shape of a slit. They reported that at such a low intensity statistically independent photons passing through a Fabry-Perot interferometer did not form the interference pattern. However, when the intensity of photons was so large that photons emitted by a lamp were in correlated states, the interference pattern was quite distinguishable.

\end{document}